\documentclass[useAMS,usenatbib,usegraphicx]{mn2e}
\newcommand*{\wn}{cm$^{-1}$}
\usepackage{verbatim}
\usepackage{hyperref}
\usepackage{xfrac}

\title[H$_2$ absorption lines in white dwarf photospheres]
  {H$_2$ Lyman and Werner band lines and their sensitivity for a variation of the proton-electron mass ratio in the gravitational potential of white dwarfs
}

\author[E.~J.~Salumbides et al.]
{E.~J.~Salumbides,$^{1,2}$\thanks{e.j.salumbides@vu.nl}
  J.~Bagdonaite,$^1$
  H.~Abgrall,$^{3,4}$
  E. Roueff,$^{3,4}$
  W. Ubachs$^1$\thanks{w.m.g.ubachs@vu.nl}\\
$^1$Department of Physics and Astronomy, LaserLaB, VU University,
  De Boelelaan 1081, 1081 HV Amsterdam, The Netherlands\\
$^2$Department of Physics and Astronomy, University of San Carlos,
  Cebu City 6000, The Philippines\\
$^3$LERMA, Observatoire de Paris, PSL Research University, CNRS, UMR8112, F-92190 Meudon, France\\
$^4$Sorbonne Universit\'{e}s, UPMC Univ. Paris 06, UMR8112, LERMA, F-75005, Paris, France\\
}

\def\LaTeX{L\kern-.36em\raise.3ex\hbox{a}\kern-.15em
    T\kern-.1667em\lower.7ex\hbox{E}\kern-.125emX}

\begin{document}

\label{firstpage}

\maketitle

\begin{abstract}

Recently an accurate analysis of absorption spectra of molecular hydrogen, observed with the Cosmic Origins Spectrograph aboard the Hubble Space Telescope, in the photosphere of white dwarf stars GD133 and GD29-38 was published in a Letter [Phys. Rev. Lett. 113, 123002 (2014)], yielding a constraint on a possible dependence of the proton-electron mass ratio on a gravitational field of strength 10,000 times that at the Earth's surface. In the present paper further details of that study are presented, in particular a re-evaluation of the spectrum of the $B^1\Sigma_u^+ - X^1\Sigma_g^+ (v',v'')$ Lyman bands relevant for the prevailing temperatures (12,000 - 14,000 K) of the photospheres. 
An emphasis is on the calculation of so-called $K_i$-coefficients, that represent the sensitivity of each individual line to a possible change in the proton-electron mass ratio. Such calculations were performed by semi-empirical methods and by \emph{ab initio} methods providing accurate and consistent values. A full listing is provided for the molecular physics data on the Lyman bands (wavelengths $\lambda_i$, line oscillator strengths $f_i$, radiative damping rates $\Gamma_i$, and sensitivity coefficients $K_i$) as required for the analyses of H$_2$-spectra in hot dwarf stars.
A similar listing of the molecular physics parameters for the $C^1\Pi_u - X^1\Sigma_g^+ (v',v'')$ Werner bands is provided for future use in the analysis of white dwarf spectra.

\end{abstract}

\begin{keywords}
 white dwarfs: molecules: H$_2$ --
 white dwarfs: photospheres --
 stars: individual: GD133, GD29-38
\end{keywords}

\section{Introduction}

Molecular hydrogen is the most abundant molecular species in the universe, but its spectrum is not so well suited
for direct observations, because the strong dipole-allowed transitions in the Lyman ($B^1\Sigma_u^+ - X^1\Sigma_g^+$) and Werner
($C^1\Pi_u - X^1\Sigma_g^+$) systems fall in the range of the vacuum ultraviolet (VUV), hence at wavelengths for which the
Earth's atmosphere is opaque.
The first observation of molecular hydrogen in space was reported in 1970~\citep{Carruthers1970} using a rocket borne spectrometer observing from high altitudes, therewith evading atmospheric absorption of the VUV radiation.
Subsequently, absorption lines in the VUV-spectrum of molecular hydrogen were observed with the {\em Copernicus} satellite telescope~\citep{Morton1976}.
Later the VUV-absorption spectrum of H$_2$ was observed with the \emph{Far Ultraviolet Spectroscopic Explorer}~\citep{Moos2000} and the \emph{Hubble Space Telescope}~\citep{Meyer2001}, all satellite-based instruments.

Spectra of molecular hydrogen in highly redshifted objects, e.g. in the line of sight of quasars, can be observed with ground-based telescopes provided the redshift of the absorbing cloud is at $z>2$.
Early observations from Earth with tentative assignment of molecular hydrogen were performed by~\citet{Levshakov1985}. Surveys of Damped-Lyman-$\alpha$ systems in search for pronounced H$_2$ absorption features were conducted by~\citet{Ledoux2003}.
The molecular hydrogen spectra in highly redshifted objects were analyzed and compared with classical~\citep{Abgrall1993a,Abgrall1993b,Abgrall1993c} and laser-based laboratory investigations of the Lyman bands~\citep{Ubachs2004,Philip2004,Hollenstein2006}, initially resulting in some indications for a possible variation of the proton-electron mass ratio ($\mu=m_p/m_e$) at cosmological redshifts~\citep{Ivanchik2005,Reinhold2006}.
Further detailed and accurate studies of molecular hydrogen at high redshift resulted in constraints on a varying proton-electron mass ratio~\citep{King2008,Malec2010,Weerdenburg2011} now yielding an overall limit of $|\Delta\mu/\mu|< 10^{-5}$ ~\citep{Bagdonaite2014a}.

Fundamental physics theories predict variation of fundamental constants on cosmological temporal and spatial scales, but the couplings between light scalar fields causing those variations may also generate dependencies of fundamental constants on local gravitational fields~\citep{Magueijo2002}. The latter phenomena were explored recently in search of a possible dependence of the fine structure constant $\alpha$~\citep{Berengut2013} and the proton-electron mass ratio $\mu$~\citep{Bagdonaite2014b} in the strong gravitational fields prevailing in the photosphere of white dwarf stars. The search for a $\mu$-dependence on a gravitational potential as strong as 10$^4$ times that at the Earth's surface was based on H$_2$ Lyman band absorptions in white dwarfs GD133 and GD29-38 initially discovered by~\citet{Xu2013,Xu2014}, and reanalyzed by~\citet{Bagdonaite2014b}.

For an analysis of the H$_2$ absorption spectrum in the hot environment of white dwarf photospheres, at typical temperatures of 12,000 - 14,000 K, accurate molecular physics data are required on excitations from highly excited ro-vibrational states in the Lyman band system. It is the purpose of the present paper to collect and re-evaluate the information on laboratory wavelengths $\lambda_i$ for the relevant Lyman bands, and perform a calculation on sensitivity coefficients $K_i$ to $\mu$-variation for the relevant lines. Together with existing information on line oscillator strengths for the transitions $f_i$ and natural lifetime parameters $\Gamma_i$ for the upper states~\citep{Abgrall2000} a constraining value for  $|\Delta\mu/\mu|$ can be derived in the presence of the strong gravitational fields near the surface of white dwarfs. In section~\ref{Moldat} an evaluation of the wavelengths $\lambda_i$ will be performed, which will be listed alongside with values for $f_i$ and $\Gamma_i$. In section~\ref{Ksec} a calculation of sensitivity coefficients $K_i$ will be presented, following different methods, one based on the semi-empirical analysis of mass dependencies of level energies, and one based on an \emph{ab initio} calculation. The combined molecular physics information on H$_2$ will be applied in section~\ref{WD} for the analysis of white dwarf spectra of GD29-38 and GD133, observed with the Cosmic Origins Spectrograph on the Hubble Space Telescope (COS-HST).

\section{Molecular data on the Lyman Bands}
\label{Moldat}

\begin{figure}
\centering
\includegraphics[width=85mm]{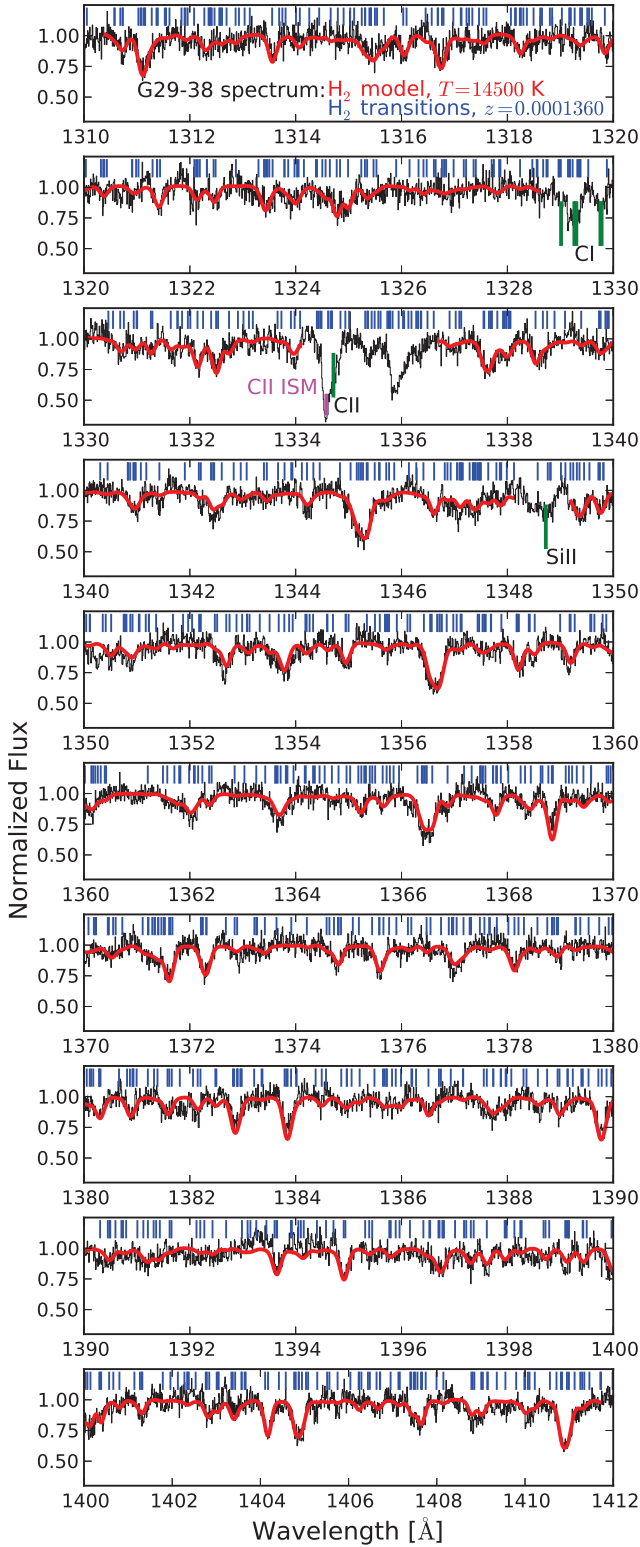}
\caption{
Spectrum of the white dwarf GD29-38 in the range 1310 - 1412 \AA\ with a model spectrum (calculated for a temperature of 14,500 K) overlaid by the (red) solid curve. The contributing H$_2$ lines in the Lyman bands are indicated by the (blue) vertical sticks. The regions dominated by CI lines (from the white dwarf photosphere) and CII lines (from the interstellar medium) are left out of the model calculations.
}
\label{WD-29-38}
\end{figure}

\begin{figure}
\centering
\includegraphics[width=85mm]{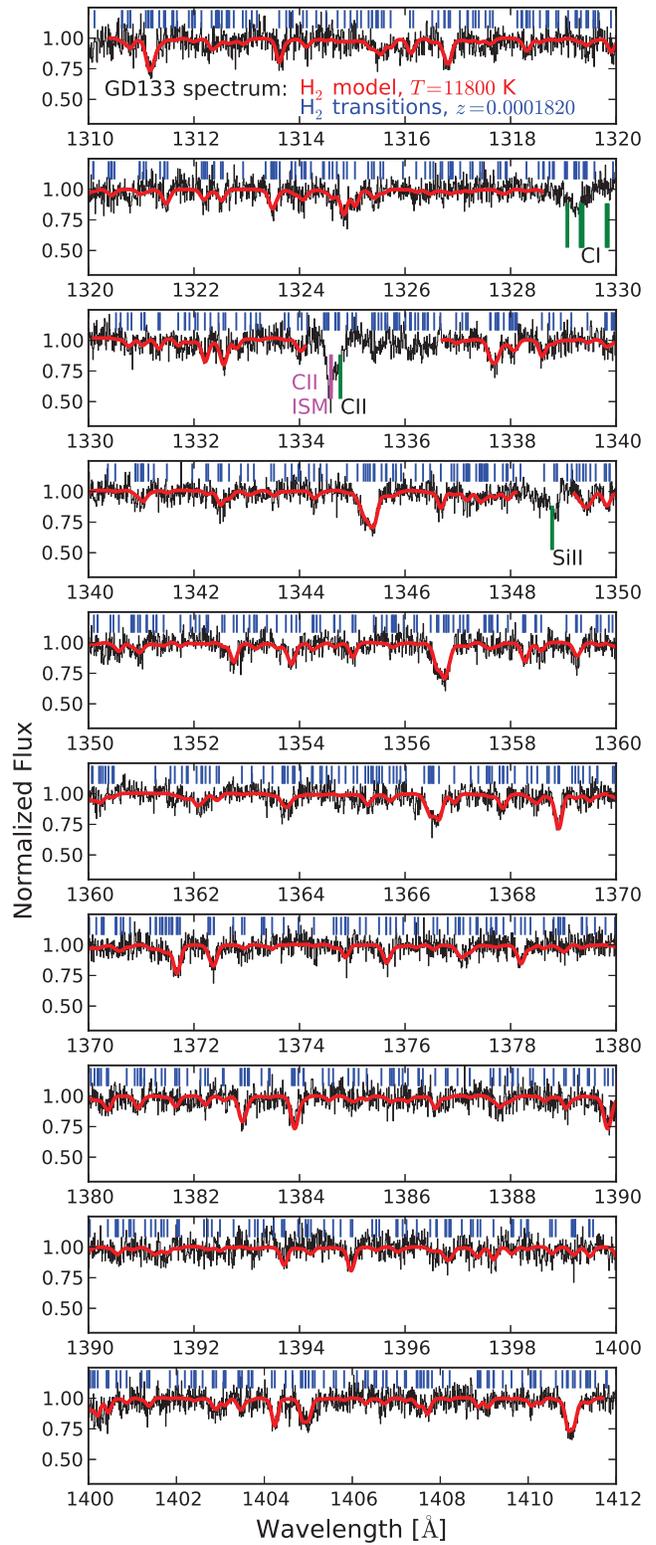}
\caption{
Spectrum of the white dwarf GD133 in the range 1310 - 1412 \AA\ with a model spectrum (calculated for a temperature of 11,800 K) overlaid by the (red) solid curve.
}
\label{WD-133}
\end{figure}

Parts of the absorption spectra of the photospheres of white dwarfs GD29-38 (WD2327+049) and GD133 (WD1116+026) obtained with COS-HST are plotted in Figs.~\ref{WD-29-38} and \ref{WD-133}.
The observed spectra~\citep{Xu2013,Xu2014} extend over the range 1144 - 1444 \AA,
with $B^{1}\Sigma _{u}^{+}$\thinspace --\thinspace $X^{1}\Sigma _{g}^{+}$ Lyman
band lines unambiguously assignable in the window 1298\thinspace --\thinspace
1444\thinspace \AA, and tentative indications of weak 
$C^{1}\Pi_{u}$\thinspace --\thinspace $X^{1}\Sigma _{g}^{+}$ Werner
band transitions in the range 1144\thinspace --\thinspace
1290\thinspace \AA. The spectra reflect the high temperatures in the white dwarf photospheres,
by the multiple vibrationally -- and rotationally
-- excited levels of the ground electronic state.
The most intense lines belong to the $B (v'=0) - X (v''=3,4)$ bands of the Lyman system.

At the high temperatures in white dwarf photospheres populations of vibrational levels up to $v''=6$ must be considered. Based on the calculations of Franck-Condon factors~\citep{Spindler1969} it can be deduced what bands in the Lyman system produce most of the absorption intensity. Whereas for excitation from $v''=0-1$ a sequence of vibrational bands contributes to the absorption, in case of excitation from $v''=3$ the Franck-Condon factors yield most of the strength for excitation to $v'=0-1$, while in case of excitation from $v''=4-5$ excitation to a single $v'=0$ level outweighs all other contributions by over an order of magnitude. This pattern is reflected in the calculations of individual line oscillator strengths $f_i$ for the Lyman bands as performed by~\citet{Abgrall2000}. Similarly, high rotational angular momentum states are populated, peaking at $J=9$ for temperatures of 12,000 K, while transitions originating from states up to $J=23$ are found~\citep{Bagdonaite2014b} to contribute to the absorption in the spectra shown in Figs.~\ref{WD-29-38} and \ref{WD-133}.

In order to model the spectra at the highest accuracy possible the wavelengths $\lambda_i$ of the relevant Lyman absorption lines have to be included in a model. The most accurate wavelengths follow from listed level energies for excited $B^1\Sigma_u^+, v, J$ and ground $X^1\Sigma_g^+, v, J$ quantum states. Starting from level energies is the superior approach over using listed experimental wavelengths, since those level energy values represent averages over several well-calibrated spectral lines. This approach was followed in the two-step determination of $B^1\Sigma_u^+, v, J$ level energies~\citep{Salumbides2008}, first measuring anchor line energies of some intermediate states in H$_2$~\citep{Hannemann2006}. A subsequent study of transitions between excited states, e.g. the $EF^1\Sigma_g^+ - B^1\Sigma_u^+$ system, provides the high accuracy on $B^1\Sigma_u^+$ level energies from the large data base of lines measured in a broad wavelength interval, including the infrared. These combined procedures resulted in level energies for $B^1\Sigma_u^+, v=0-2$ up to $J=12$, for $B^1\Sigma_u^+, v=3$ up to $J=9$, and for $B^1\Sigma_u^+, v=4-6$ up to $J=6$ at accuracies generally at 0.001 \wn\ or better~\citep{Bailly2010}. These precision experiments do not extend to very high rotational quantum numbers, needed for the analysis of the white dwarf spectra. For those higher rotational states the most accurate values are those of~\citet{Abgrall1993a}, where level energies are determined from analysis of thousands of lines observed in Lyman band emissions obtained from a low-pressure discharge~\citep{Abgrall1993c}. The values of $B^1\Sigma_u^+, v=0-6$ level energies were calculated up to $J=28$ in an \emph{ab initio} model in which the input potential curves were slightly adapted to represent the experimental data at an accuracy of 0.1 cm$^{-1}$. These calculations were further tested in a VUV-laser-based experiment performed in a plasma in which high ro-vibrational levels in the $B-X$ system were probed, yielding agreement within the estimated uncertainties of a few 0.1 \wn~\citep{Gabriel2009}.

The comparison of the level energies from~\citet{Bailly2010} and ~\citet{Abgrall1993c} for low $J$ quantum numbers yield good agreement, but with a noticeable systematic offset that is still within the estimated uncertainty of the latter. Based on this comparison, an energy correction of between 0.04-0.06 cm$^{-1}$ was applied for each vibrational series in the dataset of~\citet{Abgrall1993c}. The combined dataset comprising the low $J$ level values from~\citet{Bailly2010} and the (corrected) high $J$ level energies from ~\citet{Abgrall1993c} was employed in the calculation of the transition wavelengths for the analysis.
Similar procedures were also applied to the level energies of the $C^1\Pi_u$ state where the low $J$ level energies were taken from~\citet{Bailly2010}. For the high $J$ levels, energy corrections of between 0.04-0.13 cm$^{-1}$ was applied for the vibrational series of the $C^1\Pi_u^{+}$ states, and 0.06-0.11 cm$^{-1}$ for the $C^1\Pi_u^{-}$ states from ~\citet{Abgrall1993c}.

For the ground state one may rely on the full-fledged \emph{ab initio} calculations of all 302 bound $X^1\Sigma_g^+, v, J$ quantum levels in the ground state of H$_2$. Those values have been calculated, including Born-Oppenheimer energies, adiabatic and non-adiabatic corrections as well as relativistic and quantum-electrodynamic contributions by~\citet{Komasa2011}. Their accuracy at 0.001 cm$^{-1}$ has been tested in a number of precision experimental studies on the ground state level energies~\citep{Salumbides2011,Tan2014,Kassi2014,Niu2014} well verifying the claimed accuracy.

Wavelengths for lines in the $B^1\Sigma_u^+ - X^1\Sigma_g^+, (v',v'')$ Lyman bands are calculated from these data sets on level energies for excited and ground states. Values for the wavelengths and their accuracies, varying between 0.001 cm$^{-1}$ and 0.1 cm$^{-1}$ (on a frequency/energy scale), are listed for vibrational levels $v''=0-4$ and $v'=0-7$ in the Supplementary Material,%
\footnote{Supplementary Material is provided on the molecular physics properties of over 3000 H$_2$ Lyman lines, about a third is used in subsequent analysis of  white dwarf absorption spectra. Similar lists for Werner transitions are also provided.}
while a subset of the most intense lines is listed in Table~\ref{Lyman}.
The corresponding wavelengths for Werner transitions are also listed in Supplementary Material with a few of the most intense lines listed in Table~\ref{Werner}.
The tables contain the molecular physics knowledge of the Lyman lines to be observed in the hot environments of white dwarf photospheres; wavelengths $\lambda_i$, line oscillator strengths $f_i$, natural line broadening coefficients $\Gamma_i$~\citep{Abgrall2000}, and the sensitivity coefficients $K_i$ to be discussed in the next section.

\begin{figure}
\centering
\includegraphics[width=80mm]{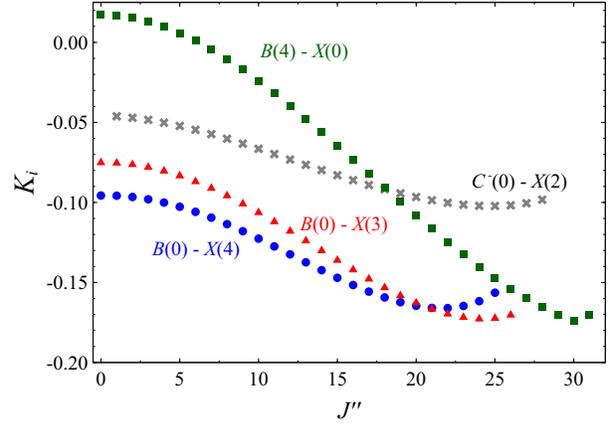}
\caption{
Calculated $K_i$-coefficients of some intense lines in the $R-$branch of H$_2$ Lyman bands.
The (green) squares represent the lines in the $B-X (v',0)$ band typically observed
in the cold environment of high redshift objects toward quasars (although only for $J=0-5$),
while (red) triangles and (blue) circles represent $B-X (0,v'')$ bands (most intense for $v''=3,4$) are typically observed in the
hot photospheres of white dwarfs.
$K_i$-coefficients for most intense Werner Q-branch lines at $T=13000$ K, belonging the $C^- - X (0,2)$ bands are also shown.
}
\label{K-coefs}
\end{figure}

\begin{table*}
\begin{minipage}{126mm}
\caption{Molecular data on H$_2$ transitions observed in spectra of hot white dwarf photospheres.
Specified are vibrational quantum numbers in excited $B, v^{\prime }$ and ground states $X, v^{\prime \prime }$, specific lines in the P and R branches,
wavelengths $\lambda_i$ in \AA , with uncertainties in between parentheses given in units of last digit;
sensitivity coefficients obtained from \emph{ab initio} $K_i^{a}$ and semi-empirical $K_i^{e}$ calculations (dimensionless); line oscillator strengths $f_i$ (dimensionless); and natural broadening coefficients $\Gamma_i$ in s$^{-1}$. 
The full listing is included in a Supplementary document. 
}
\label{Lyman}%
\begin{tabular}{c@{\hspace{10pt}}c@{\hspace{10pt}}c@{\hspace{10pt}}r@{.}l@{\hspace{10pt}}l@{\hspace{10pt}}l@{\hspace{10pt}}c@{\hspace{10pt}}c}
\hline
$B, v^{\prime }$ & $X, v^{\prime \prime }$ & Line &
\multicolumn{2}{c}{$\lambda_i$(\AA)} & $K_i^{e}$ & $K_i^{a}$ & $f_{i}$  & $\Gamma_{i} (10^{9}\, \mathrm{s^{-1}})$  \\
\hline

0	&3	&R(9)	&1\,313&376\,434\,(6)	&-0.1062	&-0.1061	&0.0494	& 1.07 \\
0	&3	&P(9)	&1\,324&595\,010\,(6)	&-0.1146	&-0.1143	&0.0504 & 1.18 \\
0	&3	&R(11)	&1\,331&963\,484\,(6)	&-0.1199	&-0.1199	&0.0472 & 0.98 \\
0	&3	&P(11)	&1\,345&177\,888\,(6)	&-0.1293	&-0.1291	&0.0508 & 1.07 \\
0	&3	&P(13)	&1\,368&662\,949\,(7)	&-0.1450	&-0.1450	&0.0513 & 0.98 \\
0	&4	&R(7)   &1\,356&487\,606\,(7)	&-0.1140	&-0.1137	&0.0821 & 1.18 \\
0	&4	&P(7)	&1\,366&395\,252\,(7)	&-0.1215	&-0.1211	&0.0715 & 1.30 \\
0	&4	&R(9)	&1\,371&422\,414\,(7)	&-0.1252	&-0.1251	&0.0816 & 1.07 \\
0	&4	&P(9)	&1\,383&659\,161\,(7)	&-0.1341	&-0.1338	&0.0739 & 1.18 \\
0	&4	&R(11)	&1\,389&593\,797\,(8)	&-0.1382	&-0.1382	&0.0816 & 0.98 \\
0	&4	&P(11)	&1\,403&982\,606\,(8)	&-0.1482	&-0.1480	&0.0765	& 1.07 \\
0	&4	&R(13)	&1\,410&648\,(1)	&-0.1522	&-0.1524	&0.0821 & 0.90 \\
0	&4	&P(13)	&1\,427&013\,40\,(8)	&-0.1630	&-0.1630	&0.0793 & 0.98 \\
0	&4	&R(15)	&1\,434&177\,(1)	&-0.1667	&-0.1670	&0.0834 & 0.84 \\
0	&4	&P(15)	&1\,452&354\,(1)	&-0.1781	&-0.1782	&0.0827 & 0.90 \\
0	&5	&R(11)	&1\,447&613\,(1)	&-0.1542	&-0.1542	&0.0951 & 0.98 \\
1	&3	&R(11)	&1\,310&900\,08\,(6)	&-0.1107	&-0.1114	&0.0489 & 0.98 \\
1	&3	&R(13)	&1\,332&315\,(1)	&-0.1254	&-0.1265	&0.0513 & 0.97 \\
1	&3	&R(19)	&1\,410&759\,(1)	&-0.1710	&-0.1730	&0.0580 & 0.72 \\
2	&2	&R(19)	&1\,338&350\,(1)	&-0.1471	&-0.1502	&0.0417 & 0.72 \\
                                        
\hline
\end{tabular}%
\end{minipage}
\end{table*}

\section{Sensitivity Coefficients $K_{i}$}
\label{Ksec}

Level energies of quantum states, as well as spectral lines of atoms and molecules depend on the values of two fundamental constants, the fine structure constant $\alpha$ and the proton-electron mass ratio $\mu$. Search for a variation of these fundamental constants under certain specific conditions, such as those in the early Universe or the presence of strong gravitational fields~\citep{Magueijo2002}, can be probed by the effect of a relative change of a constant, e.g. $\Delta\mu/\mu$ on the spectra of molecules~\citep{Jansen2014}. Here the focus is on the H$_2$ Lyman bands and how the vast amount of Lyman lines in the absorption spectra of white dwarf photospheres can be employed to detect a possible dependence of $\mu$ on the strong gravitational field near the surface of such an object that underwent gravitational collapse. For this purpose it is imperative to calculate the dependence of each individual line $i$ at wavelength $\lambda_i$ on the parameter $\mu$ via:
\begin{equation}
K_i=\frac{d\ln\lambda_i}{d\ln\mu}=-\frac{\mu}{E_B-E_X} \left(\frac{dE_B}{d\mu} -
\frac{dE_X}{d\mu} \right)
\label{K-def}
\end{equation}
where the $K_i$ are referred to as the sensitivity coefficients. The equation implies that all absorption lines exhibit a different sensitivity to a variation of $\mu$, connected to the dependence of $E_B(v,J)$ excited and $E_X(v,J)$ ground state levels to values of $\mu$.
This definition of $K$-coefficients is consistent with wavelength shifts between one set of data $\lambda_i^{\mathrm{z}}$ at $z$, and a set at $\lambda_i^0$, corresponding to the laboratory data:
\begin{equation}
\frac {\lambda_i^{\mathrm{z}}}{\lambda_i^0} = (1+z)(1+ \frac{\Delta \mu}{\mu}K_i)
\label{redshift-1}
\end{equation}
Here the parameter $z$ may signify a cosmological redshift (in the case of highly redshifted objects toward quasars), a Doppler shift (in case of objects in the interstellar medium within our galaxy), or a gravitational redshift (as in the case of absorptions occurring in high gravitational fields and probed from the Earth at a smaller gravitational potential).

Previous calculations of sensitivity coefficients for the H$_2$ Lyman bands have been carried out to model absorption in the cold clouds in the lines-of-sight of quasars, and were restricted to excitation from the lowest $v''=0$ vibrational level in the $X^1\Sigma_g^+$ electronic ground state to $B(v')$ levels. The $K_i$ calculations were performed using a semi-empirical method based on the experimentally-determined level energies, that were implemented in a Dunham formalism~\citep{Reinhold2006,Ubachs2007}. Alternatively an \emph{ab initio} approach was followed~\citep{Meshkov2006} yielding results in good agreement.
Here both methods of calculating $K_i$ sensitivity coefficients will be extended for the $B^1\Sigma_u^+ - X^1\Sigma_g^+, (v',v'')$ Lyman bands for a much wider parameter space involving many excited $v''>0$ levels in the ground state. The calculations were also extended to cover the $C^1\Pi_u - X^1\Sigma_g^+, (v',v'')$ Werner bands.

Resulting $K_i$-coefficients for $R-$branch transitions in some of the strongest Lyman bands observed in both white dwarfs are plotted in Fig.~\ref{K-coefs}. For comparison, the $K$-coefficients for the $B(4)-X(0)$ Lyman band, used for the analysis of $\mu$-variation in quasar absorption studies are also plotted, showing the higher sensitivity of the $B(0)-X(4)$ band despite the same $|\Delta v| = 4$. Note that for the low-$J$ values the sensitivity coefficients in $B(4)-X(0)$ have opposite sign.
The $K_i$-coefficients for $Q-$branch transitions of the strongest Werner band $C^-(0)-X(2)$ at $T=13000$ K is also shown.
In the following subsections the semi-empirical and the \emph{ab initio} approaches used will be presented in more detail, followed by a comparison of results obtained from either methods.

\subsection{Semi-empirical analysis of $K_{i}$}

The semi-empirical approach is related to the Dunham-based approach of ~\citet{Ubachs2007}, which extracted the sensitivity coefficients based on the Dunham representation of level energies.
However, in the present semi-empirical analysis, a full Dunham fitting is not carried out, instead the derivative $dE/d\mu$ is obtained from numerical partial differentiation with respect to the vibrational $v$ and rotational $J$ quantum numbers, with the relationship between derivatives expressed as
\begin{equation}
	\frac{dE}{d\mu}\Bigg|_{v,J} = -\frac{1}{2\mu} (v + 1/2) \frac{\partial E}{\partial v}\Bigg|_{v,J} - \frac{1}{\mu} \frac{J(J+1)}{2J+1} \frac{\partial E}{\partial J}\Bigg|_{v,J}.
\end{equation}
This results in a more direct procedure that only requires derivatives on the level energy series in the calculation of the sensitivity coefficients $K_i$.
The present approach is actually a generalization of the Dunham-based approach that is not limited to the first-order mass dependence as was considered in ~\citet{Ubachs2007}.
In practice, the Dunham fitting of the level energies is not very robust since it requires one to carefully set the number of Dunham matrix coefficients in the representation to minimize the correlations between in the fitted parameters, a complication that is bypassed in the present method.
A disadvantage of the semi-empirical approach is that local perturbations due to level energy crossings of excited states need to be treated for each individual case.
It turns out that even without accounting for such local effects, the accuracy obtained for the $K_i$ is sufficient for the present analysis.

The derivatives $dE/d\mu$ are calculated separately for the excited electronic states, $B\,^1\Sigma^+_u$ and $C\,^1\Pi_u$, and the ground electronic state $X\,^1\Sigma^+_g$. For the $B$ and $C$ states, the combined dataset from \citet{Bailly2010} and \citet{Abgrall1993c} is used, while the values from \citet{Komasa2011} are employed for the $X$ levels as discussed above.
For the purpose of comparison between the semi-empirical and \emph{ab initio} results for the separate electronic states, we introduce the \emph{level} sensitivity coefficient $\mu\,dE/d\mu$. The difference of \emph{level} sensitivity coefficients divided by the energy difference or transition energy between levels is then equivalent to \emph{transition} sensitivity coefficient $K_i$ via Eq.~(\ref{K-def}).
Whereas $K_i$ coefficients are dimensionless, level sensitivity coefficients are in units of cm$^{-1}$ and represent the corresponding shifts in the level energy if $\mu$ would change by 100\%.

\subsection{Ab initio calculation of $K_{i}$}

Contrary to previous calculations on the Lyman and Werner band systems of H$_2$ \citep{Abgrall1993a,Abgrall1993b,Gabriel2009}, where some empirical adjustments on the potential curves had been performed
to obtain a good match between experimental and computed transition wave numbers, we use here a pure {\it{ab initio}} approach which allows to introduce the $\mu$-dependence of the different quantities on a pure theoretical basis. 
We compute separately the energies of the upper (or lower) states of the electronic transitions for 11 different values of the reduced mass $\mu_{\mathrm{red}}$ of the system, varied in steps of $\Delta\mu_\mathrm{red} = 10^{-4}$ in atomic units centered around $\mu_{\mathrm{red},0} = 0.5\mu$.
These calculations are performed for each $J$ total angular momenta of up to $J =  31$.
The reference value for H$_2$ is $\mu_{\mathrm{red},0} = 918.0764$ \citep{Staszewska2002}. 

As implied in the definition of Eq.~(\ref{K-def}), we separate the effect of the electronic or nuclear (vibrational and rotational) degrees of freedom in the mass sensitivity of the level energies.
The nuclear effect is treated in the Schr{\"o}dinger equation solution for rovibrational level energies, for different reduced masses, given an electronic potential with the appropriate centrifugal repulsive term.
These sensitivities are by far the dominant contributions to $K_i$.
To improve the accuracy of the sensitivity coefficients, a partial treatment of additional electronic effects is applied by incorporating non-Born-Oppenheimer contributions, for example to the ground state electronic potential.
Non-adiabatic interaction effects in the excited electronic (BO) potentials are included in the treatment, which may significantly alter the sensitivity of specific quantum levels, in cases of level crossings.

The ground state $X$ level energies are obtained from the solution of the one-dimensional  Schr{\"o}dinger  nuclear equation where the accurate Born-Oppenheimer potential energy computed by \cite{Pachucki2010} is used.
The additional non-adiabatic contribution is derived from the computations of \cite{Komasa2011}.
The upper level energies involved in the Lyman and Werner transitions in H$_2$  are obtained from the solution of  coupled Schr{\"o}dinger  nuclear equations involving the corresponding excited electronic potentials.  The theory was described in \cite{Abgrall1987} when only $B\,^1\Sigma^+_u$ and $C\,^1\Pi_u$
rotational coupling was introduced and extended subsequently to the four interacting potentials $B\,^1\Sigma^+_u$,  $B'\,^1\Sigma^+_u$, $C\,^1\Pi_u$ and $D\,^1\Pi_u$ \citep{Abgrall1994}.
We used thus the accurate Born-Oppenheimer potential energy curves and adiabatic corrections of $^1\Sigma_u$ and $^1\Pi_u$ states of H$_2$  computed respectively by \cite{Staszewska2002} and \cite{Wolniewicz2003}.
The nonadiabatic couplings between these various potentials are taken from \cite{Wolniewicz2006}.
We have taken great care in the numerical resolution   of the coupled Schr\"{o}dinger equations  
performed with the Johnson's method \citep{Johnson1978} to avoid any spurious effect due to insufficiently small step size and insufficiently large integration length. 

The level sensitivity coefficient $\mu\,dE/d\mu$ for an upper (or lower) level is obtained from the slope of the calculated level energy for different $\mu_\mathrm{red}$ values. Subsequently, \emph{ab initio} $K_i$ values are derived from the difference of level sensitivity coefficients scaled by the corresponding transition energies according to Eq.~(\ref{K-def}).

\subsection{Comparison between both methods}

\begin{figure}
\centering
\includegraphics[width=80mm]{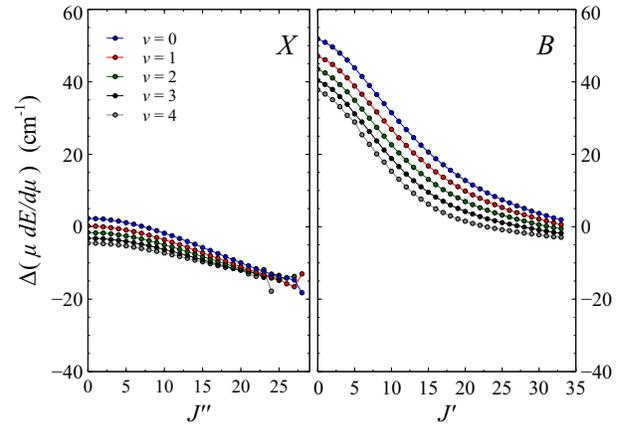}
\caption{
Difference between level sensitivity coefficients, $\Delta(\mu \frac{dE}{d\mu})$, obtained from the \emph{ab initio} and semi-empirical methods for the $v=0-5$ levels of $B$ and $X$ electronic states.
}
\label{Comparison_muDer}
\end{figure}

\begin{figure}
\centering
\includegraphics[width=80mm]{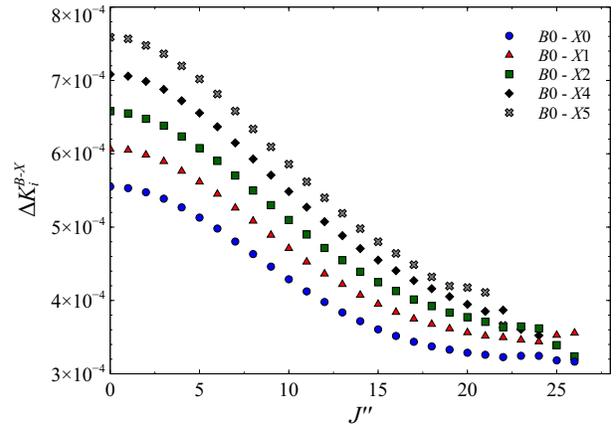}
\caption{
Difference $\Delta K_i^{B-X}$ in the sensitivity coefficients of $B-X$ transitions coefficients obtained from \emph{ab initio} and semi-empirical methods for $P-$branch transitions of the $v_B=0, v_X=0-5$ bands.
}
\label{ComparisonKmu}
\end{figure}

In order to verify the consistency of the methods to derive sensitivity coefficients and to assess
their uncertainty a comparison was made between the two distinct methods of calculation.
For this purpose we have applied the semi-empirical method on the \emph{ab initio}  $X$ and $B$ state level energies, calculated for the most accurate value of $\mu=1\,836.152\,672\,45\,(75)$ as listed in CODATA \citep{CODATA2010}.
The difference of the level energy sensitivities, $\Delta(\mu dE/d\mu)$, for the $B$ and $X$ electronic states, obtained using the semi-empirical and \emph{ab initio} methods is plotted in Fig.~\ref{Comparison_muDer} for the first few $v_{B,X}$ quantum numbers.
In the $X$ state comparison, some discrepancy is observed at higher $J$ that may be attributed to the numerical limitation of the semi-empirical method at the boundary where there are fewer values of $J$ belonging to states with increasing $v$ quantum numbers.
The results of both methods are in better agreement for the ground state $X$ than those of electronically excited levels, for example $B$.
This may be attributed to the isolation of the ground state to any other interacting states, since the agreement for the $C\,^{1}\Pi^-_u$ state that has interactions with fewer states, is comparable to that of the $X$ state.
It turns out that the deviations when using experimental values of the level energies for the $B$ state or the more accurate level energies for the ground state from \citet{Komasa2011} are not significantly different to the trends observed here.

The transition sensitivity coefficients are calculated using Eq.~(\ref{K-def}), where the difference of the level energy sensitivities, $\mu\,(dE_{B,X}/d\mu)$, is scaled by corresponding the transition energy.
A comparison of the difference $\Delta K_i^{B-X}$ in transition sensitivity coefficients is shown in Fig.~\ref{ComparisonKmu} for $P$-branch transitions for $(v_B=0, v_X=0-5)$ bands. For the transitions observed in white dwarf spectra described below, the \emph{ab initio} and semi-empirical methods are in agreement to a few $10^{-4}$ which is more than sufficient for the present application. 
Similar procedures were applied to the $C\,^1\Pi_u$ state to calculate the sensitivity coefficients for the Werner transitions, where the results from both methods also show good agreement.

It is noted that the deviations between the two methods discussed above only lead to a few times $10^{-4}$ contribution in absolute terms to the the final $K_i$ coefficients, that is, 0.1\% in relative terms.
Despite these discrepancies being the maximum observed, we adopt this as a conservative uncertainty estimate for the $K_i$ coefficients presented here.

\begin{table*}
\begin{minipage}{126mm}
\caption{Strongest H$_2$ Werner transitions at temperatures of $T=13000$ K.
Specified are vibrational quantum numbers in excited $C, v^{\prime }$ and ground states $X, v^{\prime \prime }$,
wavelengths $\lambda_i$ in \AA , with uncertainties in between parentheses given in units of last digit;
sensitivity coefficients obtained from \emph{ab initio} $K_i^{a}$ and semi-empirical $K_i^{e}$ calculations (dimensionless); line oscillator strengths $f_i$ (dimensionless); and natural broadening coefficients $\Gamma_i$ in s$^{-1}$. 
Line transitions in of the Q branch involve the $C\,^1\Pi_u^-$ states, while P-branch transitions involve the $C\,^1\Pi_u^+$ states.
The full listing is included in a Supplementary document.
}
\label{Werner}%
\begin{tabular}{c@{\hspace{10pt}}c@{\hspace{10pt}}c@{\hspace{10pt}}r@{.}l@{\hspace{10pt}}l@{\hspace{10pt}}l@{\hspace{10pt}}c@{\hspace{10pt}}c}
\hline
$C, v^{\prime }$ & $X, v^{\prime \prime }$ & Line &
\multicolumn{2}{c}{$\lambda_i$(\AA)} & \multicolumn{1
}{c}{$K_i^{e}$} & $K_i^{a}$ & $f_{i}$  & $\Gamma_{i} (10^{9}\, \mathrm{s^{-1}})$  \\
\hline
0	&1	&Q(19)	&1\,142&712\,(1)	&-0.0898	&-0.0900	&0.0735       & 1.09 \\
0	&2	&Q(13)	&1\,143&692\,(1)	&-0.0798	&-0.0799	&0.0767       & 1.13 \\
0	&2	&P(13)	&1\,151&641\,(1)	&-0.0900	&-0.0912	&0.0399       & 1.13 \\
0	&1	&P(19)	&1\,152&290\,(1)	&-0.0992	&-0.0987	&0.0492       & 1.10 \\
0	&2	&Q(15)	&1\,156&072\,(1)	&-0.0861	&-0.0863	&0.0789       & 1.12 \\
0	&1	&Q(21)	&1\,158&244\,(1)	&-0.0963	&-0.0965	&0.0761       & 1.07 \\
0	&2	&P(15)	&1\,164&871\,(1)	&-0.0953	&-0.0962	&0.0396       & 1.12 \\
0	&1	&P(21)	&1\,168&414\,(1)	&-0.1051 	&-0.1055	&0.0515       & 1.08 \\
0	&2	&Q(17)	&1\,169&257\,(1)	&-0.0918	&-0.0920	&0.0814       & 1.10 \\
0	&1	&P(23)	&1\,184&811\,(1)	&-0.1094	&-0.1099	&0.0553       & 1.06 \\
                                                
\hline
\end{tabular}%
\end{minipage}
\end{table*}

\section{Application to White Dwarf Spectra}
\label{WD}

The sensitivity coefficients, both from \emph{ab initio} $K_i^a$ and semi-empirical $K_i^e$ methods,  for the most intense lines are included in Table~\ref{Lyman} for the Lyman bands.
The $K_i^e$ entries were calculated using the accurate level energies from \citet{Bailly2010} and \citet{Abgrall1993c}, with applied corrections as discussed above, for the electronically excited states, while the most accurate level energy calculations of \citet{Komasa2011} were used for the ground state levels. 
A full list that also includes the Werner bands is presented in the Supplementary Material.
In the spectral range between 1310 and 1420 \AA\ used in the analysis, the strongest Werner transitions are at least an order of magnitude weaker than the strongest Lyman lines.
The most intense Werner lines for H$_2$ at temperatures of $T=13000$ K lie in the 1140 to 1250 \AA\ wavelength range, with the first few listed in Table~\ref{Werner}.
The Q-branch transitions involve levels of the $C\, ^{1}\Pi^{-}_u$ state, while the P- and R-branches involve the $C\, ^{1}\Pi^{+}_u$ state.
The comprehensive database is produced for the purpose of future analyses of H$_2$-absorption in high-temperature white dwarf photospheres.

\begin{figure}
\centering
\includegraphics[width=80mm]{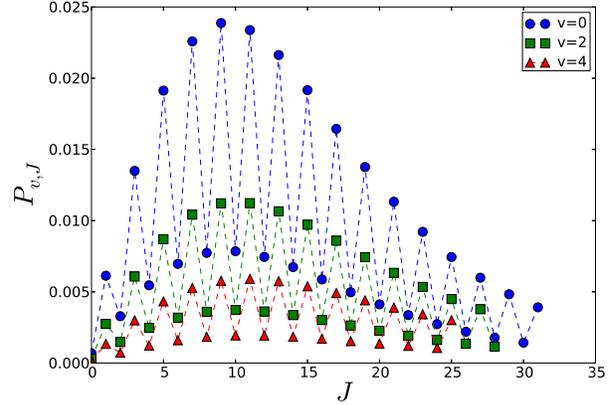}
\caption{
Population of $v=0,2,$ and $4$ levels in the electronic ground state $X$ of H$_2$ for $T=13000$ K.}
\label{Population}
\end{figure}

In Tables~\ref{Lyman}~and~\ref{Werner}, the molecular physics information on the H$_2$ Lyman and Werner bands is re-evaluated and tabulated in terms of accurate wavelengths $\lambda_{v',v'',J',J''}^B$ and $\lambda_{v',v'',J',J''}^C$, line oscillator strengths $f_{v',v'',J',J''}^{BX}$ and $f_{v',v'',J',J''}^{CX}$, excited state damping factors $\Gamma_{v',J'}^B$ and $\Gamma_{v',J'}^C$, and sensitivity coefficients $K_{v',v'',J',J''}^{BX}$ and $K_{v',v'',J',J''}^{CX}$. Note that single-prime indices refer to excited states and double-prime to ground state levels. This information can be used to produce a model function of the absorption spectrum at a certain temperature, under the assumption of thermodynamic equilibrium.
For the purpose of producing a model spectrum a partition function for ground state level energies is calculated, for a certain temperature $T$ with
\begin{equation}
P_{v,J}(T) = \frac{g_I(J)\, (2J+1) \exp{\left(\frac{-E_{v,J}}{kT}\right)}}{\sum\limits_{v=0}^{v_{\mathrm{max}}} \sum\limits_{J=0}^{J_{\mathrm{max}}(v)} g_I(J)\, (2J+1) \exp{\left(\frac{-E_{v,J}}{kT}\right)} }
\label{partition-function}
\end{equation}
where $k$ is the Boltzmann constant and $g_I(J) = (2 - (-1)^J)$ is a nuclear-spin degeneracy factor, the ortho:para ratio that leads to a 3:1 ratio for odd:even levels.
Unlike cold interstellar absorption clouds, vibrational and rotationally-excited levels of the $X$ ground state of H$_2$ are substantially populated in hot white dwarf photospheres as shown in Fig.~\ref{Population}.

The intensities of the absorption lines is then given by:
\begin{equation}
I_i=N_{\mathrm{column}} f_{v',v'',J',J''} P_{v'',J''}(T)
\end{equation}
where $f_i$ are the line oscillator strengths specific for a certain transition and $N_{\mathrm{column}}$ is the total
column density of H$_2$ absorbers, summed over all quantum states, populated with a fraction $P_{v'',J''}(T)$. Here we restrict ourselves to the $B-X$ Lyman bands as we analyze the spectral region where these are unambiguously detected.
The spectra recorded from the photospheres of the white dwarfs are modeled by taking the wavelength positions $\lambda_i$, assigning them an intensity $I_i$ and a Gaussian width $b$ (the same for each line) convolved with the Lorentzian broadening coefficient $\Gamma_i$, resulting in a Voigt profile. Finally the spectrum is allowed to undergo a redshift factor $(1+z)$.
The model spectrum is treated by the software package {\sc vpfit} 
\footnote{{\sc vpfit} developed by R. F. Carswell et al.; available at \url{http://www.ast.cam.ac.uk/~rfc/vpfit.html} }
where parameters $N_{\mathrm{column}}$, $b_i$ and $z$ are optimized in a fit to the entire data set, for each white dwarf spectrum.
The fits are run for various input temperatures, hence starting from different partition functions, and the fit results for the $T$-value with a minimum $\chi^2$ are listed in Table~\ref{Tab-WD}. The redshift $z$ is mainly the result of the gravitational potential.
The listed $b$-parameter is the result from the fit, deconvolved from an instrument function, corresponding to 17 km/s. As such it can be interpreted as the Doppler width in the absorbing photospheres, where it is noted that a Maxwellian distribution at 12000 K would give a Doppler width of $\sim 10$ km/s.

Finally for the optimum value of the temperature a possible variation of the $\mu$ constant was included via Eq.~(\ref{redshift-1}). Resulting constraints on a dependence of the proton-electron mass ratio are also listed in Table~\ref{Tab-WD}, in terms of a relative value $\Delta\mu/\mu$.
The fitting of the spectra involved both the updated values of the sensitivity coefficients, and a refined treatment of the normalization of the temperature-dependent populations.
The present fitting results, e.g. for the column densities and temperatures, are then slightly different from those reported in \citet{Bagdonaite2014b}. 
Verifications of the latter procedure were done for a number of temperatures, and no significant deviations found on $\Delta\mu/\mu$. For completeness the value of the gravitational potential
$\phi =GM/Rc^{2}$, with $G$ Newton's constant, $R$ and $M$ the radius and mass of the white dwarf, and $c$ the speed of light, are included in the Table. These values may be compared to the gravitational potential at the Earth's surface $\phi = 9.8 \times 10^{-9}$, which is in fact produced by the gravitational field of the Sun.
Hence, as a general result, the analysis of the white dwarf spectra constrains the proton-electron mass ratio to $|\Delta\mu/\mu| < 5 \times 10^{-5}$ at gravitational potentials $\phi = 10^4 \phi_{\mathrm{Earth}}$.

\begin{table}
\begin{center}
\caption{Resulting parameters from the analysis of the H$_2$ absorption spectrum for the two white dwarfs GD133 and GD29-38 analyzed.}
\label{Tab-WD}
\begin{tabular}{lcc}
\hline \hline
 Parameter  & GD133   & GD29-38 \\
\hline
 $\log N_{\mathrm{column}}$\,[cm$^{-2}$]  &  $15.817 \pm 0.007$             & $15.959 \pm 0.005$ \\
 $T$\,[K]                                 &  $11100 \pm 470$                & $13500 \pm 340$     \\
 $b$\,[km/s]                              &  $14.50 \pm 0.58$               & $18.29 \pm 0.42$   \\
 $z$                                      &  0.0001819(11)                  &  0.0001358(8)   \\
 $\Delta\mu/\mu$                          &  $(-2.3 \pm 4.7)\times 10^{-5}$ & $(-5.8 \pm 3.7)\times 10^{-5}$ \\
 $\phi_{\mathrm{WD}}$                     &  $1.2 \times 10^{-4}$           & $1.9 \times 10^{-4}$ \\
 \hline
\end{tabular}
\end{center}
\end{table}

The use of H$_2$ Lyman band lines for constraining $\Delta\mu/\mu$ may be mimicked by offsets in the wavelength calibration of the astronomical spectra, in particular in cases where a correlation exists between wavelength and sensitivity coefficients $K_i$. This possibility was discussed in the context of possible cosmological variation of $\mu$, or limitations thereof, as probed through H$_2$ absorption in the colder environments of high redshift galaxies, where typical excitation temperatures of 50 K prevail~\citep{Ledoux2003,Ivanchik2005,Malec2010,Weerdenburg2011,Bagdonaite2014a}. Under such conditions only a few rotational states, $J=0-5$, are populated for a single $v''=0$ vibrational level, and the $K_i$-coefficients for the Lyman bands strongly correlate with wavelength~\citep{Ubachs2007}. For those studies the simultaneous use of Lyman and Werner band absorptions lifts the correlation somewhat, as shown in Fig. 4 in \citet{Ubachs2007}.

\begin{figure}
\centering
\includegraphics[width=85mm]{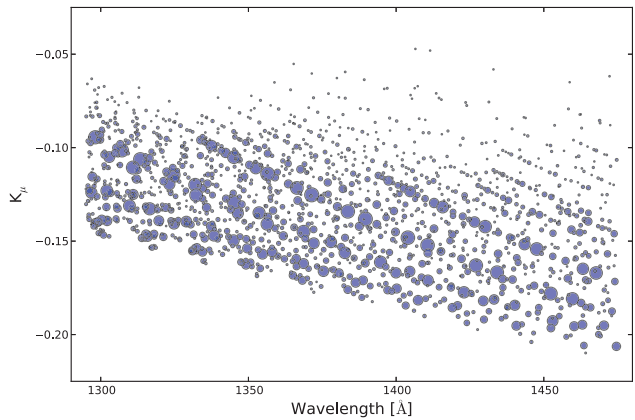}
\caption{
Calculated $K_i$-coefficients plotted vs. wavelength with relative line intensities
as they appear in the spectra of the photospheres of white dwarfs marked as the sizes of the points.
The line intensities refer to the product of line oscillator strengths $f_{v',v'',J',J''}^{BX}$ and the fractional population $P_{v,J}(T)$ at a temperature of 12,000 K.
}
\label{K-coeffs-WD}
\end{figure}

For the present case of the Lyman absorption bands at 12000 K the situation is much different as exemplified in Fig.~\ref{K-coeffs-WD} showing the calculated $K$-coefficients plotted as a function of wavelength, while the size of the data points reflects the (relative) line intensities of the specific lines in the spectrum. This figure demonstrates that the correlation between $K$ and $\lambda$ is lifted -- at each wavelength in the spectrum various lines with differing $K_i$ contribute to the absorption spectrum. This feature makes the derivation of constraints on $\Delta\mu/\mu$ more robust.

\section{Conclusion}

In this study the molecular physics information on the $B^1\Sigma_u^+ - X^1\Sigma_g^+ \, (v',v'')$ Lyman bands and $C^1\Pi_u - X^1\Sigma_g^+ \, (v',v'')$ Werner bands of the H$_2$ molecule is collected and re-evaluated resulting in a listing of wavelengths $\lambda_i$, line oscillator strengths $f_i$, and radiative damping factors $\Gamma_i$, while sensitivity coefficients $K_i$ for a possible change of the proton-electron mass ratio $\mu$ are calculated. This database is implemented for the analysis of H$_2$ absorption spectra in the photospheres of GD133 and GD29-38 white dwarfs stars yielding a constraint on a possible dependence of the proton-electron mass ratio of $\Delta\mu/\mu < 5 \times 10^{-5}$ in a gravitational potential of $\phi = 10^4 \phi_{\mathrm{Earth}}$. The study focuses on the Lyman bands for $v'=0-3$ and $v''=0-6$
, corresponding to the strongest H$_2$ absorption features, observed with the Cosmic Origins Spectrograph aboard the Hubble Space Telescope.

Future spectroscopic studies on similar hot photosphere white dwarfs may cover the shorter wavelength region where
$C^1\Pi_u - X^1\Sigma_g^+ \, (v',v'')$ Werner lines contribute. For that purpose calculations on the molecular physics data on the Werner bands are tabulated in the present study. For improving the analysis methods extended data would be required along the following lines: (i) The laboratory wavelengths $\lambda_i$ for Lyman and Werner lines involving rotational quantum states $J>12$ are only accurate to 0.1 \wn; laser-based studies could improve those values by two orders of magnitude, and it was shown by~\citep{Salumbides2011} that laser-based studies on very high $J$-states are feasible. (ii) If white dwarf photospheres with enhanced column densities can be detected, spectroscopic features of the HD molecule should become observable. For a detailed analysis of such environments the molecular physics properties of the Lyman and Werner bands of HD should then be investigated, along the lines as pursued by~\citet{Ivanov2008,Ivanov2010}.

\section*{acknowledgements}
The authors acknowledge support from the Netherlands Foundation for Fundamental Research of Matter (FOM).
We have used data from the Cycle 18 -- Program 12290 (PI M. Jura) observations with the Hubble Space Telescope.

\label{lastpage}

\end{document}